# Title: A major asymmetric dust trap in a transition disk

**Authors:** Nienke van der Marel[1], Ewine F. van Dishoeck[1,2], Simon Bruderer[2], Til Birnstiel[3], Paola Pinilla[4], Cornelis P. Dullemond[4], Tim A. van Kempen[1,5], Markus Schmalzl[1], Joanna M. Brown[3], Gregory J. Herczeg[6], Geoffrey S. Mathews[1], Vincent Geers[7]

**Affiliations:**

[1] Leiden Observatory, Leiden University, P.O. Box 9513, 2300 RA Leiden, The Netherlands

[2] Max-Planck-Institut für Extraterrestrische Physik, Giessenbachstrasse 1, 85748 Garching, Germany

[3] Harvard-Smithsonian Center for Astrophysics, 60 Garden Street, Cambridge, MA 02138, USA

[4] Heidelberg University, Center for Astronomy, Institute for Theoretical Astrophysics, Albert Ueberle Str. 2, 69120, Heidelberg, Germany

[5] Joint ALMA Offices, Av. Alonso de Cordova, Santiago, Chile

[6] Kavli Institute for Astronomy and Astrophysics, Peking University, Yi He Yuan Lu 5, Hai Dian Qu, 100871 Beijing, PR China

[7] Dublin Institute for Advanced Studies, 31 Fitzwilliam Place, Dublin 2, Ireland

**Abstract**: The statistics of discovered exoplanets suggest that planets form efficiently. However, there are fundamental unsolved problems, such as excessive inward drift of particles in protoplanetary disks during planet formation. Recent theories invoke dust traps to overcome this problem. We report the detection of a dust trap in the disk around the star Oph IRS 48 using observations from the Atacama Large Millimeter/submillimeter Array (ALMA). The 0.44-millimeter-wavelength continuum map shows high-contrast crescent-shaped emission on one side of the star originating from millimeter-sized grains, whereas both the mid-infrared image (micrometer-sized dust) and the gas traced by the carbon monoxide 6-5 rotational line suggest rings centered on the star. The difference in distribution of big grains versus small grains/gas can be modeled with a vortex-shaped dust trap triggered by a companion.

**Main Text:** While the ubiquity of planets is confirmed almost daily by detections of new exoplanets *(1)*, the exact formation mechanism of planetary systems in disks of gas and dust around young stars remains a long-standing problem in astrophysics *(2)*. In the standard core-accretion picture, dust grains must grow from submicron sizes to ~10 $M_{Earth}$ rocky cores within the ~10 Myr lifetime of the circumstellar disk. However, this growth process is stymied by what is usually called the "radial drift and fragmentation barrier": Particles of intermediate size (~1 m at 1 AU, or ~1 mm at 50 AU from the star) acquire high drift velocities toward the star with respect to the gas *(3,4)*. This leads to two major problems for further growth *(5)*: First, high-velocity collisions between particles with different drift velocities cause fragmentation. Second, even if particles avoid this fragmentation, they will rapidly drift inward and thus be lost into the star before they have time to grow to planetesimal size. This radial drift barrier is one of the most persistent issues in planet formation theories. A possible solution is dust trapping in so-called pressure bumps: local pressure maxima where the dust piles up. One example of such a pressure bump is an anticyclonic vortex which can trap dust particles in the azimuthal direction *(6-10)*.

Using the Atacama Large Millimeter/submillimeter Array (ALMA), we report a highly asymmetric concentration of millimeter-sized dust grains on one side of the disk of the star Oph IRS 48 in the 0.44 millimeter (685 GHz) continuum emission (Fig. 1). We argue that this can be understood in the framework of dust trapping in a large anticyclonic vortex in the disk.

The young A-type star Oph IRS 48 (distance ~ 120 pc, 1 pc = $3.1 \cdot 10^{13}$ km) has a well studied disk with a large inner cavity (deficit of dust in the inner disk region), a so-called transition disk. Mid-infrared imaging at 18.7 µm reveals a disk ring in the small dust grain (size ~50 µm) emission at an inclination of ~50°, peaking at 55 AU radius (1 AU = $1.5 \cdot 10^8$ km = distance from Earth to the Sun) or 0.46 arcseconds from the star *(11)*. Spatially resolved observations of the 4.7 µm CO line, tracing 200-1000 K gas, show a ring of emission at 30 AU radius and no warm gas in the central cavity *(12)*. This led to the proposal of a large planet clearing its orbital path as a potential cause of the central cavity. While these observations provide information on the inner disk dynamics, they do not address the bulk cold disk material accessible in the millimeter regime.

The highly asymmetric crescent-shaped dust structure revealed by the 0.44 millimeter ALMA continuum (Fig. 1) traces emission from millimeter-sized dust grains and is located between 45 and 80 AU (± 9 AU) from the star. The azimuthal extent is less than one third of the ring with no detected flux at a 3σ level (2.4 mJy/beam) in the northern part (see Fig. S1). The peak emission has a very high signal-to-noise ratio of ~390 and the contrast with the upper limit on the opposite side of the ring is at least a factor of 130. The complete absence of dust emission in the north of IRS 48 and resulting high contrast makes the crescent-shaped feature more extreme than earlier dust asymmetries *(10,13)*. The spectral slope of the millimeter fluxes (0.44 mm combined with fluxes at longer wavelengths *(14)*) is only 2.67 ± 0.25 ($F_\nu \propto \nu^\alpha$), suggesting that millimeter-sized grains *(15)* dominate the 0.44 mm continuum emission. However, the gas traced by the $^{12}$CO 6-5 line from the same ALMA data set indicates a Keplerian disk profile characteristic of a gas disk with inner cavity around the central star (Fig. 1B). $^{12}$CO 6-5 emission is detected down to a 20 AU radius, which is consistent with the hot CO ring at 30 AU *(14)*. This indicates that there is indeed still some CO inside the dust hole, with a significant drop of the gas surface density inside of ~25 ± 5 AU. The simultaneous ALMA line and continuum observations leave no doubt on the relative position of gas and dust.

The observations thus indicate that large millimeter-sized grains are distributed in an asymmetric structure, but that the small micrometer-sized grains are spread throughout the ring. To our knowledge, the only known mechanism that could generate this separation in the distribution of the large and small grains is a long-lived gas pressure bump in the radial and azimuthal direction. The reason that dust particles get trapped in pressure bumps is their drift with respect to the gas in the direction of the gas pressure gradient: $\vec{v}_{dust} - \vec{v}_{gas} \propto \vec{\nabla} p$ *(3,4)*. In protoplanetary disks without vortices this gradient typically points inward, so dust particles experience the above-mentioned rapid radial drift issue. If, however, there exists (for whatever reason) a local maximum of the gas pressure in the disk (i.e. where $\vec{\nabla} p_{gas} = 0$ and $\vec{\nabla}^2 p_{gas} < 0$), then particles would converge toward this point and remain trapped there *(3,5)*, avoiding both inward drift and destructive collisions *(14)*. Because small dust particles are strongly coupled to the gas they will be substantially less concentrated toward the pressure maximum along the azimuthal direction than large particles. Various mechanisms have been proposed that could produce a local pressure maximum in disks, for instance when there is as a "dead zone" (*16*), or a substellar companion/planet (*14,17*) in the disk, hindering accretion. Until recently, however, the presence of such dust pressure traps was purely speculative, because pre-ALMA observations did not have the spatial resolution and sensitivity necessary to constrain the distribution of gas and dust required for testing dust trapping models *(18)*.

Although vortices in models have an azimuthal gas contrast up to only a factor of a few *(16,19,20)*, models predict that even a very minor pressure variation in the *gas* ring will trap the dust efficiently, leading to a strong lopsided azimuthal asymmetry in the *dust* ring if the vortex is long-lived. A gas contrast of only 10% can create a dust contrast of 100 for large dust particles in ~$10^5$ years *(9)*, so a long-lived azimuthal pressure bump can readily explain the observed high asymmetry in the dust structure of IRS 48. Vortices created by planets have been shown to survive over at least $10^5$ years *(8)*. Even though these vortices tend to diffuse at longer timescales, ~$10^5$ years is enough time to create strong dust concentrations that remain even if the vortex disappears. It takes millions of years to even out these dust concentrations completely *(9)*. More generally, vortices are expected to be long-lived if they have an elongation (aspect ratio of arc length over width) of at least 4 *(21)*. The accumulated dust in IRS 48 has an elongation of ~3.1 (± 0.6).

We present a detailed numerical model *(14)*, showing the feasibility of our proposed scenario (Fig. 2). Given the central cavity in the Oph IRS 48 disk, we propose a substellar companion as the cause of the inner cavity which also creates a long-lived ring of enhanced pressure outside the planetary orbit. The gas densities inside the cavity are decreased with the level depending on the companion mass and disk viscosity *(22,23)*. The drop of the gas surface density at ~25 ± 5 AU in combination with the shape and steep radial drop of the millimeter dust emission at 45 AU suggests that this substellar companion is located between the star and dust trap around ~20 AU, and has a mass of at least 10 $M_{Jup}$ *(17)*. The presented model with these parameters shows that the radial overdensity at the edge of the cavity is Rossby unstable, leading to the production of an anti-cyclonic vortex *(14)*. Dust accumulation in this pressure bump results in the spatial separation between the gas and the millimeter dust emission *(17,24)*. Other than the hole, the high gas velocities are symmetric in the east and west and consistent with Keplerian motion around a 2 $M_{Sun}$ star (see Fig. 1). Any gas density variation along the azimuthal direction cannot be observed directly in the CO 6-5 observations, due to high line optical depth within the disk and foreground absorption, but $C^{17}O$ data are not inconsistent with a full gas ring *(14)*.

Regardless of the formation mechanism, the ALMA observations clearly show a concentration of dust grains within a small region of the disk. The total measured dust emission corresponds to 9 Earth masses assuming a dust temperature of 60 K. The millimeter observations confirm dust growth up to a maximum grain size of $a_{max} = 4$ mm. Including larger grains, the dust mass could be a factor of $\sqrt{a_{max}/4}$ [mm] larger *(14)*. The mass is similar to the full disk dust masses found in other young disks *(25)*. This amount of large dust in a small area will favour grain growth up to ~ meter size until the fragmentation barrier *(5)*. Further growth to planetesimal sizes is possible when additional mechanisms such as the sweeping-up of small particles by larger seeds and bouncing effects including mass-transfer are considered *(26,27)*. Since these closely formed planetesimals will scatter and disperse along the ring on short time scales, it is not possible to continue growth and form a planetary core with regular orderly growth models within ~10 Myr at this large distance from the star. On the other hand, the dust trap could initiate the formation of a Kuiper Belt around IRS 48 such as that found in our own Solar System at comparable radii *(28)*. The dust trap as a 'Kuiper Belt Object factory' is analogous to a 'planet factory' at smaller radii around other stars, where both the fragmentation barrier is higher and the collisional growth is faster *(5)*. Thus, the possibility of dust trapping as the start of core formation could help to explain the observations of massive planets at smaller radii around A-stars such as found in HR 8799 *(29)* and beta Pictoris *(30)*.

Dust asymmetries have been hinted at in other disks by SubMillimeter Array (SMA) observations *(13)*, and are clearly seen in earlier ALMA observations *(10)*. The low image fidelity of the SMA data (low sensitivity and spatial sampling) and the lower contrast leave room for multiple interpretations, although a relation to vortices has been hinted at *(10)*. In contrast, the ALMA observations of IRS 48 with their unprecedented spatial resolution and sensitivity show a contrast of at least 130 in the continuum along the ring, with no indications of a highly asymmetric small dust/gas distribution. Alternative scenarios are discussed to be less likely *(14)*. A long-lived azimuthal gas pressure bump triggered by a companion, followed by particle-trapping, appears to be the most viable scenario that could produce this. The key feature is the observed separation between big and small dust grains/gas, which is a direct consequence of the dust-trapping model.

**Acknowledgments:** We thank M. Benisty and W. Lyra for useful discussions. This paper makes use of the following ALMA data: ADS/JAO.ALMA# 2011.0.00635.SSB. ALMA is a partnership of ESO (representing its member states), NSF (USA) and NINS (Japan), together with NRC (Canada) and NSC and ASIAA (Taiwan), in cooperation with the Republic of Chile. The Joint ALMA Observatory is operated by ESO, AUI/NRAO and NAOJ. The data presented here are archived at http://www.alma-allegro.nl/science and the full project data (2011.0.00635.SSB) will be publicly available at the ALMA Science Data Archive, https://almascience.nrao.edu/alma-data/archive.

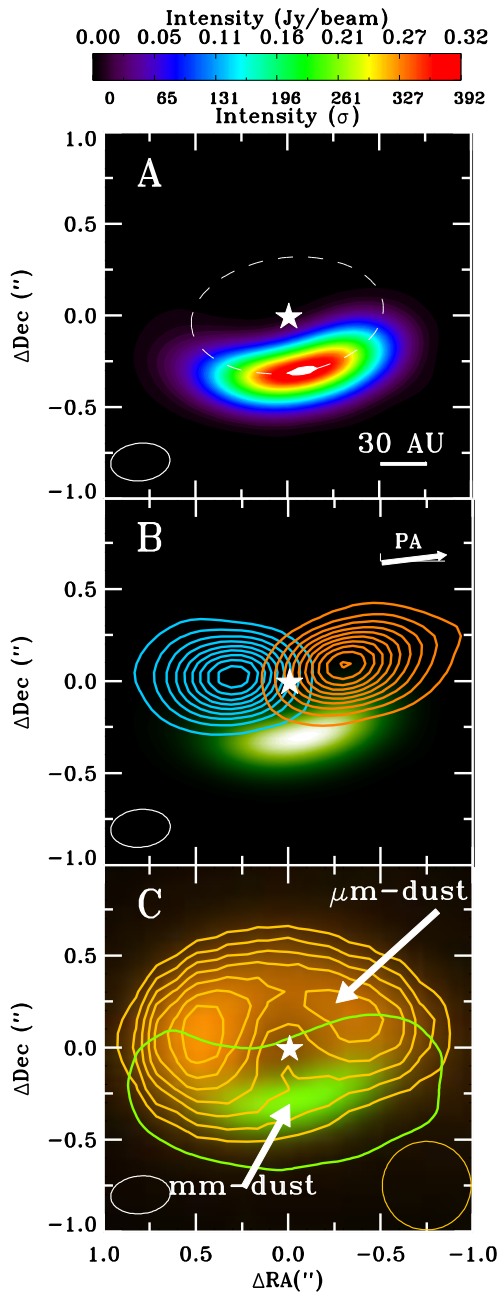

**Fig. 1. IRS 48 dust and gas observations.** The inclined disk around IRS 48 as observed with ALMA Band 9 observations, centered on the star (white star symbol). The ALMA beam during the observations is 0.32" × 0.21" and is indicated with a white ellipse in the lower left corner. A) the 0.44 millimeter (685 GHz) continuum emission expressed both in flux density and relative to the rms level ($\sigma$ = 0.82 mJy/beam). The 63 AU radius is indicated by a dashed ellipse. **B)** The integrated CO 6-5 emission over the highest velocities in contours (6,12,...,60$\sigma_{CO}$ levels, $\sigma_{CO}$ = 0.34 Jy km s$^{-1}$): integrated over -3 to 0.8 km s$^{-1}$ (blue) and 8.3 to 12 km s$^{-1}$ (red), showing a symmetric gas disk with Keplerian rotation at an inclination $i = 50°$. The green background shows the 0.44 millimeter continuum. The position angle is indicated in the upper right corner.

**C)** The VLT VISIR 18.7 μm emission in orange contours (36-120$\sigma_{VISIR}$ levels in steps of 12$\sigma_{VISIR}$, $\sigma_{VISIR}$ = 0.2 Jy arcsec$^{-2}$) and orange colors, overlayed on the 0.44 millimeter continuum in green colors and the 5$\sigma$ contour line in green. The VISIR beam size is 0.48" in diameter and is indicated with an orange circle in the bottom right corner.

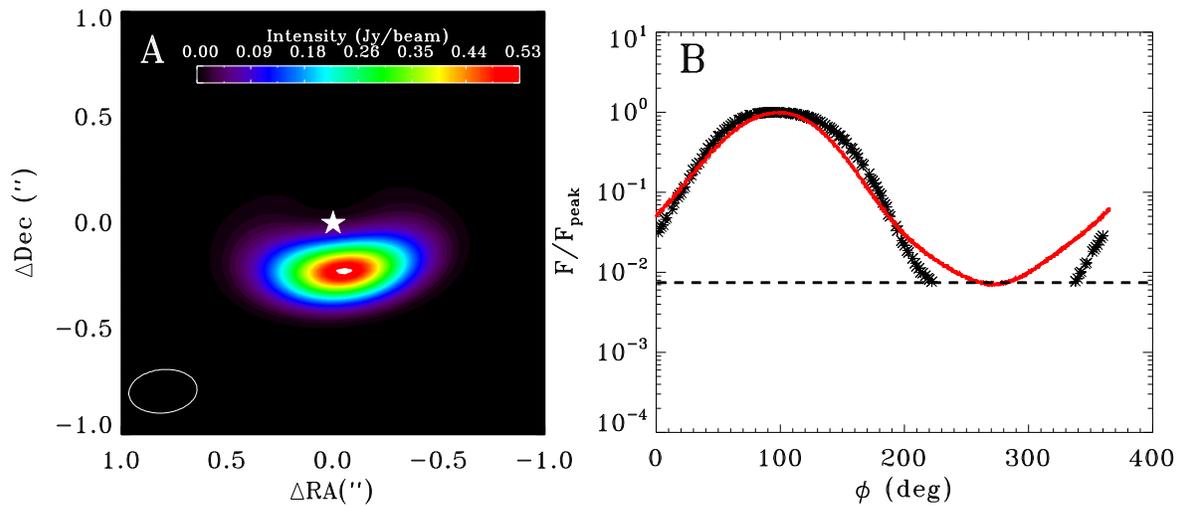

**Fig. 2. Model of the dust asymmetry. A)** Simulated ALMA Band 9 dust continuum images at 0.44 millimeter (685 GHz) for our model using the dust trapping scenario *(9)*. Details of the model and model parameters are derived based on the IRS 48 properties *(14)*. The high azimuthal contrast is similar to that found in the data. **B)** Normalized logarithmic azimuthal cut at the peak emission radius through the dust ring of the observations (black) and the ALMA-simulated model (red), showing the large contrast in the mm dust density between the maximum and the opposite side of the ring. The (normalized) $3\sigma$ upper limit of the ALMA data is indicated with a dashed line. The data points below the normalized $3\sigma$ level have been removed for clarity.

**Supplementary Materials:**

www.sciencemag.org
Materials and Methods
Figs. S1, S2, S3, S4, S5
References (31-54)

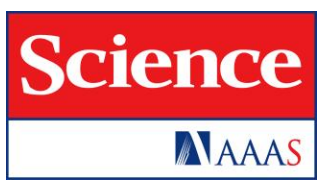

Supplementary Materials for

A major asymmetric dust trap in a transition disk

Nienke van der Marel, Ewine F. van Dishoeck, Simon Bruderer, Til Birnstiel, Paola Pinilla, Cornelis P. Dullemond, Tim A. van Kempen, Markus Schmalzl, Joanna M. Brown, Gregory J. Herczeg, Geoffrey S. Mathews, Vincent Geers

.

correspondence to: nmarel@strw.leidenuniv.nl

**This PDF file includes:**

Materials and Methods

Figs. S1 to S5

**Materials and Methods**

IRS48

The A0 star IRS 48 a.k.a. WLY 2-48 (J2000 RA=16 27 37.18, Dec=-24 30 35.3) is located in the ρ Ophiuchi star formation region at a distance of 120 parsec *(31)*. A ring-like structure peaking at ~55 AU radius was found by spatially resolved 18.7 μm imaging of the dust continuum with the Very Large Telescope (VLT)-VISIR instrument *(11)*. The 18.7 μm emission is dominated by the thermal dust emission of small grains (~50 μm) and is merely tracing the surface layer of the disk which is irradiated directly by the star, since thermal dust emission is very optically thick at this wavelength *(15)*. The 18.7 μm image appears to be brighter in the north (Fig 1C, *11*). This is due to the inclination of the disk, with our line of sight directly to the inner wall. The near (southern) part of the inner wall is hidden by the intervening body of the disk itself *(32)*. The bright peak at 55 AU radius implies a large scale height of the disk. The inner radius of the small grain disk is not constrained by the VISIR image, as the inner 20-55 AU may sit in the shadow of an inner disk and may thus not be heated enough to emit brightly at 18.7 μm.

Interferometric imaging with the SubMillimeter Array (SMA) in the very extended configuration at 0.85 mm (345 GHz) revealed a single elliptical structure, that could be modeled with a hole size of only 13 AU *(33)*. Without the simultaneous dust and gas observations available with ALMA, the SMA dust continuum was modeled as if centered on the star rather than offset. Since the gas line kinematics allow a very precise determination of the stellar position, the combined ALMA observations of line and continuum within the same dataset is crucial for accurate interpretation. The crescent-shaped structure is expected to be similar at 345 GHz: when the continuum as found in the ALMA observations is convolved with the beam from the SMA observations, the result is indeed the same shape as found in the SMA image.

Previous information on the gas in the IRS 48 disk comes from VLT-CRIRES spectra of the CO ro-vibrational transition at 4.7 μm clearly showing extended CO emission *(12)*. Modeling of its extent and position is compatible with a hot gas ring at ~30 AU radius in Keplerian rotation around the young A star. However, the lack of hot CO emission inside 30 AU may be due to a lack of excitation of the vibrationally excited level and does not imply a lack of colder gas inside the hole, which could exist due to shielding of an inner disk. Given the fact that polycyclic aromatic hydrocarbons (PAHs) are still abundant inside the dust hole *(11)*, a complete absence of molecular gas would be unlikely, especially since accretion is ongoing *(34)*.

ALMA observations

The ALMA observations of IRS 48 were taken using Band 9 in Early Science Cycle 0 using the extended configuration. Data were taken on June 6th and July 17th 2012 in three execution blocks of 1.7 hours each (one on June 6th and two on July 17th). Between these three execution blocks, 18 to 21 antennas were used with baselines up to 390 meter. The average precipitable water vapor (PWV) levels during the executions were 0.50, 0.34 and 0.17 mm respectively. The spectral setup consisted of four spectral windows, centered at 674.00 , 678.84, 691.47 and 693.88 GHz. This setup was meant to target the $C^{17}O$ $J$=6-5, CN $J$=6-5, $^{12}CO$ $J$=6-5, and $H^{13}CO^+$ $J$=8-7 transitions. Each spectral window consists of 3840 channels, a channel separation of 488 kHz and thus a bandwidth of 1875 MHz. The resulting velocity resolution is 0.21 km s$^{-1}$ using a reference of 690 GHz. The astrometric accuracy of the ALMA data is measured to be on average ~30 milliarcseconds.

Reduction and calibration of the data was done using the Common Astronomy Software Application (CASA) version 3.4. The bandpass was calibrated using quasar 3c279 and fluxes were calibrated against Titan. For Titan, the flux calibration was done using only the shortest ~30% of the baselines, since Titan was resolved out at longer baselines. For the second execution block taken on July 17th, the original bandpass and flux calibrators were quasar J1924-292 and asteroid Juno respectively. Both sources are significantly weaker than 3c279 and Titan. As such, after thorough comparison of the calibration solutions of the first execution block, as well the applications on the gain calibrators over both executions taken that night, the solutions of 3c279 and Titan were preferred over the solutions of J1924-292 and Juno.

The gains were calibrated using at first the solutions of quasars J1625-254 and J1733-130, which were observed at 12 minute intervals. It was found that using just J1625-254 resulted in the best solution. After applying the gain solutions, it was found that the resulting amplitudes and phases of IRS 48 were dominated by short-term (10 minutes or less) phase fluctuations. The calibration was improved by calibrating these fluctuations on the continuum of IRS 48 itself. This resulted in a net gain of 40% in S/N. As the continuum of IRS 48 is resolved, the unselfcalibrated continuum solution served as the model for this self-calibration. No changes in shape larger than 5% were seen as a result of this self-calibration. In the end the overall calibration uncertainty is less than 20%.

The continuum emission from IRS 48 was extracted from the total coverage by combining the line-free channels from all four spectral windows, flagging the edge channels and the channels containing the strong CO 6-5 emission. No other areas containing lines were flagged as no other strong lines were identified. The total bandwidth of the continuum is ~6.5 GHz, with a mean frequency of 684.55 GHz.

The image conversion of the continuum and line data was done using natural weighting, providing a synthesized beam of 0.32" × 0.21" at a position angle of 96°. For the cleaning of the continuum an elliptical cleaning mask around the continuum shape was used, derived from the unselfcalibrated continuum shape, padded by half a beam at the 3σ level of the unselfcalibrated image. For the lines, a rectangular cleaning mask centered on the detected emission peaks was used was used, extending 0.24" in all directions to cover most assumed emission. This mask was applied and changed adaptively to the emission of each channel and each peak with emission larger than 3σ in the various iterations of cleaning of the dataset. For the weaker lines, the final mask of the $^{12}$CO cleaning was used. The RMS noise level is 0.82 mJy/beam for the continuum and ~18 mJy/beam per 1 km s$^{-1}$ bin for the line data. This noise level is measured in a region between 2 and 5" from the central source position, avoiding any contamination from the disk emission. The continuum shows isotropic noise levels in all areas of the primary beam and no systematic effects. The total continuum flux is 950 mJy. To confirm that the dust trap is not an artifact of the reduction steps, different sizes and shapes of clean masks were compared to the result seen in the main paper. This was done both on the selfcalibrated and the unselfcalibrated image, with no structural differences seen. From the comparison we conclude that no more than 2% of the presented noise level can be attributed to the choice of calibration methods, much lower than the overall calibration uncertainty.

Morphology of the continuum emission

The extent of the dust trap was estimated by a simple surface density profile of the dust emission, assuming a fourth power Gaussian-like distribution in both radial and azimuthal direction:

$$\Sigma(R) = \Sigma_R e^{\frac{-(R-R_c)^4}{2R_w^4}}$$

$$\Sigma(\phi) = \Sigma_\phi e^{\frac{-(\phi-\phi_c)^4}{2\phi_w^4}}$$

with $R_w$ and $R_c$ the half-width and center in the radial direction and $\phi_w$ and $\phi_c$ the half-width and center in azimuthal direction. The full profile $\Sigma(R,\phi) = \Sigma(R) \cdot \Sigma(\phi)$ and the normalization constant product $\Sigma_R \cdot \Sigma_\phi$ is chosen such as to match the scale of the observations. These high power profiles are required to match the steep rise in both radial and azimuthal direction and broad peak in the data. Although the arc is only just resolved in the radial direction, it can not be infinitesimally thin considering the high total dust continuum flux of 950 mJy and a typical dust temperature of 60 K. The temperature is based on a generic radiative transfer calculation of a transition disk with a dust hole around an A-type star. Applying the inclination of $i=50°$, position angle of 97° (measured from the CO data; North to East) and the central position (derived from the CO Keplerian model described below) and convolving the total surface density model with the ALMA beam from the observations, the best fit to the continuum data was found for $R_c = 63$ ($\pm$ 8) AU, $R_w = 18$ ($\pm$ 2) AU, $\phi_c = 95$ ($\pm$ 10)° and $\phi_w = 50$ ($\pm$ 5)°. The main uncertainty arises from the uncertainties in the stellar position (~0.05"). Defining the extent of the profile as $2 \cdot R_w$ and $2 \cdot \phi_w$, the dust extends from ~45 - 80 ($\pm$ 9) AU and covers in the azimuthal direction an arc length of ~100°, less than a third of the disk ring. As discussed in the main text, these characteristics of the dust emission make this source very different from what has been observed before *(10,13,18)*.

The elongation (aspect ratio of arc length over width) of an arc is defined as:

$$e = \frac{R_c \phi_w [\text{rad}]}{R_w}$$

The elongation of our dust trap is ~3.1 ($\pm$ 0.6) with the derived parameters. If the dust trap is entirely coincident with the gas vortex, this elongation indicates whether the vortex is long-lived: in general vortices are expected to be long-lived if they have an elongation of at least 4 *(21)*.

The high north/south contrast is shown in Fig. S1; although a narrow 2σ emission (σ = 0.82 mJy/beam) arc indicates some dust emission in the north, this is not sufficient to measure the contrast. However, this image demonstrates that the emission in the south (390σ peak emission) is at least a factor 130 higher than in the north.

The radial structure of this disk is as follows (see Fig. S2): the gas shows a surface density decrease around ~25 $\pm$ 5 AU, whereas the dust inner rim is at least 45 AU away from the star.

The model presented later in the Supplement confines the location of the companion to 15-20 AU, although it is possible that multiple companions are present. While the large dust grains are azimuthally concentrated, the smaller micrometer-sized grains traced by 18.7 μm continuum imaging extend over the full ring *(10)*. The small grains could be either following the gas, only depleted in the inner most region by the companion *(35)*, or be distributed along a dust ring around 55 AU.

The ALMA continuum data were overlayed on the VISIR 18.7 μm (Fig. 1C) by aligning their central positions, determined by the CO Keplerian velocity profile for the ALMA line data and by the VISIR PAH images (taken to be centered on the star) for the 18.7 μm image. The 18.7 μm image and PAH images were taken in sequence so the relative alignment errors are small.

Dust mass

The continuum flux of the dust emission can be written as

$$F_\nu \sim \Omega \times B_\nu(T_{dust}) \times (1 - e^{\tau_\nu})$$

with solid angle $\Omega = A/d^2$ with $A$ emission area, $d$ distance and $\tau_\nu$ the opacity. The area of the continuum emission seen in the ALMA images is $A = \pi((R_c + R_w)^2 - (R_c - R_w)^2) \times 2\phi_w/360° \times \cos(i) = 2544$ AU$^2$, using the estimates derived in the previous section. The dust temperature is estimated to be ~ 60 K at a radius of 60 AU. For the total flux density of 950 mJy at 685 GHz, we obtain $\tau_{685\,GHz} = 0.43$. Flux loss due to larger scale emission was measured to be less than 20% for the 230 and 345 GHz emission for spatial scales >100 AU *(33)*. For the ALMA data, the spatial filtering occurs for >500 AU, so flux loss by spatial filtering will be even less than 20%.

The opacity is defined as $\tau_\nu = \rho_{dust}\,\kappa_\nu\,s$ with $\rho_{dust}$ the dust density (g cm$^{-3}$), $\kappa_\nu$ the dust mass opacity (cm$^2$ g$^{-1}$) and $s$ the length of the slab along the line of sight. Therefore the dust mass $M_{dust}$ can be calculated as:

$$M_{dust} = As\rho_{dust} = A\frac{\tau_\nu}{\kappa_\nu}$$

For 685 GHz and a dust temperature of 60 K, the Rayleigh-Jeans limit is justified and the slope of the (sub)millimeter emission $\alpha$ ($F_\nu \propto \nu^\alpha$) can be directly related to the power-law index of the dust opacity $\kappa_\nu \propto \nu^\beta$, where $\alpha = 2 + \beta$. The value of $\beta$ is estimated by making a power law fit to the (sub)millimeter fluxes at 230, 345 and 685 GHz (50, 160 and 950 mJy *(33)*, the 685 GHz flux density derived in this work) and we obtain $\beta = 0.67 \pm 0.25$. This low value of $\beta << \beta_{ISM} \approx 1.7$ is indicative of grain growth *(15)*. Since dust grains larger than ~ 3λ do not contribute to the opacity, dust growth can be confirmed up to a grain size of at least 4 millimeter due to the detected dust emission at 1.3 mm.

For a dust-size distribution $N(a) \sim a^{-3.5}$ and amorphous silicate spheres, the dust mass opacity is approximately *(15)* $\kappa_{685\,GHz} = 9/\sqrt{a_{max}[mm]}$ cm$^2$ g$^{-1}$ (valid for a$_{max} \geq 1$ mm). We thus obtain a dust mass of

$$M_{dust} = 4.5 \times \sqrt{a_{max}[mm]}\,M_{Earth}$$

With $a_{max} > 4$ mm this yields a lower limit of the dust mass of 9 $M_{Earth}$. This mass does not include larger bodies (boulders or planetesimals).

CO 6-5 lines

The $^{12}$CO 6-5 line ($E_U = 116$ K) is detected throughout most of the disk of IRS 48 with an rms level of 47 mJy in 0.21 km s$^{-1}$ velocity channels and a peak S/N of ~45. Fig. S3 shows the flux profile over the entire region of the disk. The profile shows a strong absorption feature around 3.5 km s$^{-1}$, caused by foreground clouds also seen in C$^{18}$O 3-2 observations *(36)*. The Local Standard of Rest source velocity is derived from the symmetry of the $^{12}$CO 6-5 profile as 4.55 km s$^{-1}$. Due to the small offset of the foreground cloud velocity compared to the disk velocity, the red-shifted emission of the disk appears to be stronger than the blue-shifted emission. Inspection of the high velocity channels shows that the profile is in fact symmetric, as expected from a Keplerian disk. Folding the full spectrum over the source velocity of 4.55 km s$^{-1}$ also shows this symmetry (Fig. S3B). Furthermore, the red velocity channels that are not contaminated by the foreground absorption extend within the region of dust continuum emission.

In order to derive the inner radius $R_{in}$ of the gas hole, the kinematics of the CO 6-5 line were studied. The fastest gas detected in the line wings agrees both kinematically and spatially with a location at 20 AU. The maximum detected velocities are -3 and 12 km s$^{-1}$ or +/- 7.5 km s$^{-1}$ from the source velocity. The emission in these channels has a separation of 0.33", in agreement with gas at 20 AU in Keplerian rotation around a 2 $M_{Sun}$ star at an inclination of 50º. The non-detection of faster gas in both the line wings and strongly decreased emission at the central position at any velocity imply a significant drop in surface density in the inner 20 AU. This is slightly interior to the hot CO gas ring at 30 AU detected at 4.7 μm *(12)*. The CRIRES observations trace much warmer gas which must by illuminated by the central star but the general kinematics are in good agreement with the gas seen by ALMA.

Due to the foreground absorption in the $^{12}$CO 6-5 line around the source velocity (Fig. S3), this emission does not reveal a full gas ring: emission is missing in the north and south part of the disk. However, the small dust grains, that are coupled to the gas, show emission both in the north and south as well as east and west (Fig. 1C, *11*). Similarly, the near infrared CRIRES lines (*12*) show hot gas emission also to the north and south of the star. Both observations are consistent with the presence of gas along the entire ring.

The $^{12}$CO 6-5 line is likely optically thick, so deviations from a symmetric full outer disk (including azimuthal density variations) can not be traced directly from these $^{12}$CO observations. Optically thin C$^{17}$O 6-5 emission shows significant detections (4-6σ, σ = 25 mJy/beam) in the eastern and western parts of the disk. In the north and south C$^{17}$O 6-5 emission is not detected, but this can be understood by limb brightening due to the inclination of the source, where emission on the sides of the disk appears up to twice as bright as in the north and south. The C$^{17}$O 6-5 observations are thus also not inconsistent with gas along the entire ring.

Integrating over the channels with significant C$^{17}$O detections (962 mJy km s$^{-1}$) and calculating the mass for an average gas temperature between 50 and 200 K, we find a total gas mass between 19 and 27 $M_{Jup}$. The integrated flux $F_{int}$ is related to the number of C$^{17}$O molecules $N_{C17O}$ by

$$F_{int} = \frac{N_{C^{17}O} f_T A_{65} hc}{4\pi d^2}$$

with $f_T = 13 \exp(-113.3 \text{ K}/T)/Q(T)$ the fraction of molecules in $J=6$, $Q(T)$ the partition function, $A_{65}$ the Einstein A coefficient of the 6-5 transition and $d$ the distance. This is converted into gas mass (assuming all is in $H_2$) by

$$M = N_{C^{17}O} m_{C^{17}O} X_{H_2/^{12}CO} X_{^{12}CO/C^{17}O}$$

with $X$ the relative abundances (2500 for $^{12}CO/C^{17}O$, $1.2 \cdot 10^4$ for $H_2/^{12}CO$) and $m_{C17O}$ the mass of a $C^{17}O$ molecule ($4.8 \cdot 10^{-23}$ g).

For comparison, the dust mass of 9 $M_{Earth}$ derived from the millimeter-sized dust suggests a gas mass of 900 $M_{Earth}$ = 2.8 $M_{Jup}$ assuming that this dust was originally smeared out over the entire disk with a standard gas/dust mass ratio of 100.

No overdensity at the central location of the dust trap is detected in the $C^{17}O$ emission, but the high dust continuum emission makes weak $C^{17}O$ emission more difficult to observe there. The large gas mass of 19-27 $M_{Jup}$ that is seen even outside the dust trap proves that there is a lot of gas along the disk ring, even without the (undetected) gas emission at the location of the dust trap. An overdensity of a factor 100 at the dust trap location compared to the east and west would make the total disk mass as high as ~1 $M_{Sun}$, which is very unrealistic, as would be the implied gas/dust ratio of $10^5$ (*18*). Moreover, such a high gas mass should have readily been detected in $C^{17}O$ even in the south. A gas density contrast within a factor of a few in the dust trap compared with the east and west can not be ruled out, however. Alternatively, if the emission in the east or west was the only location with a high gas overdensity, the dust would have been trapped at that location.

Altogether, the data can be modeled with a full gas ring with possible azimuthal density variations. The amount of variation cannot be constrained by the current data set.

Dust trapping

Trapping of dust particles has been proposed as a solution for the radial drift and fragmentation barrier. Dust trapping is a natural consequence of pressure gradients, since particles tend to drift with respect to the gas in the direction of the pressure gradient. In its simplest form one could have an axisymmetric disk with a mass accumulation at some radius, leading to a ringlike pressure trap along the radial direction with $dp_{gas}(r)/dr = 0$ and $d^2 p_{gas}(r)/dr^2 < 0$ *(3,5,37,38)*. In a radial pressure bump, where the pressure gradient is zero, the drift velocity vanishes for particles of all sizes, so that not only the drift barrier but also destructive collisions can be avoided.

The presence of a massive planet has been demonstrated to create radial pressure bumps due to a reduction of the gas surface density in the inner part of the disk *(17)*. By combining two-dimensional hydrodynamic gas simulations using FARGO *(39)* with coagulation/fragmentation dust evolution models *(40)*, it has been shown that dust grains converge toward these ring like pressure bumps and grow efficiently *(41)*. The efficiency of the dust trapping (the largest produced grain size) and the location of the dust depend on the planet orbit, planet mass and (for planet mass < 5 $M_{Jup}$) the turbulence in the disk. For planets of more than 5 $M_{Jup}$, the pressure bump and therefore also the dust can be located at distances of more than twice the star-planet separation *(17)*. The gas accumulation at the outer edge of the cavity depends on the planet mass and disk viscosity *(22)*.

If the overdensity at the outer edge of the cavity is high enough, it becomes Rossby unstable *(20,42)* and forms vortices that could be long-lived (~$10^5$ years) *(8)*. A vortex is defined as a gas region where the perturbed velocity spins around a center. In case of an overdensity created by a planet, the Rossby instability creates anticyclonic vortices (super-Keplerian motion at the inner and sub-Keplerian motion at the outer radius). Such a vortex acts as a long-lived non-axisymmetric pressure bump and can trap dust in the azimuthal direction along the ring *(9)*. The azimuthal location of the dust trap is tied to variations in the gas density rather than the planet location, because the material in the outer disk only sees the planet as an orbital ring as it is moving much faster than the outer disk. The density contrast in the dust can become significantly larger than the contrast in the gas, depending on the coupling of the dust to the gas summarized in the Stokes number:

$$St = \frac{\rho_s a}{\rho_{gas} c_s \sqrt{8/\pi}} \Omega_k$$

where $\rho_s$ is the intrinsic dust density (density of the material out of which the dust is made), $a$ the particle radius, $\rho_{gas}$ the local gas mass volume density, $c_s$ the vertical isothermal sound speed and $\Omega_k$ the Keplerian rotation rate. Particles with $St \ll 1$ are strongly coupled to the gas and move along with the gas, while particles with $St \gg 1$ move with their own velocity i.e. Keplerian velocity. In pressure maxima gas moves with Keplerian velocity, so there is no friction any more between gas and dust and as a consequence particles naturally get stuck in the pressure maximum.

The trapping is, however, never perfect, because even small amounts of turbulence will, to some extent, counteract the converging motion of the dust particles. As a result, in a steady state situation, small dust particles will be substantially less strongly concentrated toward the trapping point than large particles. In the case of IRS 48 it is possible that the small grains, traced by the 18.7 µm emission, are still radially trapped (although not azimuthally), but since the inner radius of the 18.7 µm emission can not be constrained this possibility can not be confirmed.

For estimating the mass and location of the planet in the IRS 48 disk, we use the Hill radius $r_H$, defined as $r_H = r_p (M_p/3M_*)^{1/3}$ where $r_p$ is the orbital radius of the planet. The outer edge of a gas cavity created by a planet is expected to be at most at 5 $r_H$ *(23)* whereas the location where the gas pressure bump forms - hence dust accumulates - is ~10 $r_H$ for planets with $M_p > 5$ $M_{Jup}$ *(17)*. In the case of IRS 48, with a gas hole edge at ~25 AU and a dust trap peaking at ~63 AU with the inner edge at ~45 AU, these approximate relations can not reproduce the large gas–dust separation exactly, but it indicates that the mass of the companion, likely located at 17-20 AU, has to be more massive than 10 $M_{Jup}$.

Model of the IRS 48 dust trap

A model is presented in Figure 2 reproducing the high azimuthal contrast in the dust trap in IRS 48. The gas surface density features a cavity carved out by a companion.

For the disk-planet interaction process, we use the two-dimensional hydrodynamical code FARGO *(39)*. The FARGO simulations are done considering a single planet in a fixed orbit (no migration effects are taken into account) to study the resulting gas density distribution when a massive planet carves a cavity in a disk. The used parameters are: stellar mass $M_{star} = 2$ $M_{Sun}$, companion mass $M_p = 10$ $M_{Jup}$, companion orbital radius $r_p = 20$ AU and turbulence $\alpha_t = 10^{-4}$. The initial gas surface density profile in FARGO was taken as a power-law $\Sigma_{gas} = \Sigma_0 (r/r_p)^{-1}$ with $\Sigma_0 = 1.35 \cdot 10^{-4}$ $M_{star}/r_p^2 = 3.0$ g cm$^{-2}$. This number is based on the estimated gas mass of the IRS 48 disk, assuming a gas/dust ratio of 100 with the gas distributed over the entire disk, but it turns out that the gas mass of the disk does not matter much for the nature of the vortex. For the vertical disk structure, we use an aspect ratio $h/r$ (height over radius) = 0.05 at $r_p$ and a flaring disk with a flaring index of 0.25. The radial dust temperature structure was based on a generic radiative transfer calculation of a transition disk with a dust hole around an A-type star. The gravitational potential is smoothed around the planet by $\varepsilon = 0.5$ $r_H$ ($r_H$ is the Hill radius of the planet). The inner boundary conditions are non-reflecting. FARGO uses a fixed grid in cylindrical coordinates $(r, \varphi)$. The number of pixels $n$ in each direction is $n_r \times n_\varphi = [512, 757]$. The radial grid is logarithmically spaced. The radial boundaries are $r_{in} = 0.1$ $r_p$ and $r_{out} = 7.5$ $r_p$.

These simulations show that a 10 $M_{Jup}$ planet at 20 AU creates a Rossby unstable radial pressure bump that becomes a vortex within $10^3$ year (Fig. S4). The vortex continuously evolves (there is no steady-state solution) but it does not disappear for at least $10^5$ years, which is sufficient to trap the dust efficiently. Recent work *(43)* shows that vortices in 3D simulations decay quickly compared to 2D observations when the origin of the Rossby instability disappears, but that there is no decay when the pressure gradient is sustained. Since the planet sustains the Rossby unstable overdensity, this is not an issue in our case. Therefore, the long lifetime of 2D vortices is well justified. The azimuthal gas contrast of the vortex at the outer edge of the cavity could reach values of 2-3 (Fig. S4).

The dust evolution and resulting dust distribution is calculated separately from the FARGO simulations because it is computationally expensive to do simulations of gas and dust evolution with grain growth simultaneously. We can not use the FARGO output directly because vortices are dynamical objects, whose shape varies with time. The FARGO simulations demonstrate that vortices survive long enough and have high enough gas contrast to trap particles and model the observations.

The radial gas disk profile from the FARGO simulations (Fig. S5A) is used as the initial condition for the dust evolution model. The dust evolution code *(39)* is run on this profile, simulating the growth, fragmentation and radial transport of dust to calculate the radial size distribution. For the dust evolution, we consider a grid of grain size from 1 μm to 10 cm and an intrinsic dust density $\rho_s = 0.8$ g cm$^{-3}$.

To study the azimuthal asymmetry, we now redistribute at each radius the dust using a sinusoidal perturbation *(9)* with a gas contrast of 3, as a steady-state solution based on the vortex in the FARGO simulations described above. The size distribution from the radial profile is distributed azimuthally at each radius using the analytical solution for the gas contrast *(9)*. This leads to a weak over-density of micrometer-sized grains, but a very strong overdensity of millimeter-sized grains (see Fig. S5C).

The resulting grain size distribution as function of radius and azimuth was used to calculate the opacities and the resulting continuum intensity maps, assuming low optical depth and spherical magnesium-iron-silicate grains *(44)*. The resulting continuum emission is imaged using CASA (v. 3.4.0) using the same configurations as in the actual ALMA observations.

The final image (Fig. 2A) shows that the dust trap created by this model has the same morphology and high azimuthal contrast as observed in IRS 48 (Fig. 2B). The model does not predict the exact same radial extent as the dust trap in IRS 48: the model trap is somewhat closer to the star and narrower in the radial direction. The peak emission is also higher than in the observational data. A more massive companion or placement at a larger radius would increase the radius of the dust trap, but also moves the gas hole edge too far out. A more complex system with multiple companions may solve this problem.

Alternative scenarios

The highly asymmetric dust structure in the Oph IRS 48 disk can be explained by a vortex dust trapping model (Fig. 2 and associated discussion), with the vortex formed from a pressure bump created by a substellar companion. In this section, we discuss other scenarios that could potentially lead to this kind of asymmetry and explain why they are less likely based on the available observations.

*Alternative origin of the pressure bump*

The vortex is induced by a Rossby instability of a radial pressure bump, independent of its origin. In this work we present a pressure bump generated by a gas density gradient. Although photoevaporation *(45)* could also create a cavity with an overdensity at the edge, photoevaporative clearing is unlikely in the case of IRS 48 as it still has an accretion rate of $\sim 10^{-8}$ $M_{Sun}$ yr$^{-1}$ *(34)*, much higher than typical EUV photoevaporation rates ($\sim 10^{-10}$ $M_{Sun}$ yr$^{-1}$) *(45)*. In any case, the large hole radii observed for IRS 48 put this source outside the cases that can be reproduced by photoevaporation only *(46)*.

A pressure bump could also be generated by a viscosity gradient. This gradient exists along the edge of a so-called dead zone: a region of low viscosity below the surface layer of the disk, which occurs if the gas does not have a high enough ionization fraction to enable magnetorotation instabilities that transport the gas. Dead zones form in regions where stellar X-rays and interstellar cosmic rays cannot penetrate and the temperature is not high enough to collisionally ionize the gas, as in the midplane in the inner disk *(47)*. Rossby unstable pressure bumps are formed at the edges of this dead zone *(48)* and the resulting vortices will trap the dust *(16)*. The presence of a dead zone can not be excluded from the data presented here, so a dead zone could have formed the unstable pressure bump as well. However, the absence of gas in the inner region of the disk implies a density gradient that is already sufficiently steep to become a vortex, so a dead zone is not required and we consider this origin less likely.

*Gravitational instabilities*

If a disk is very massive, gravitational instabilities *(49)* in the gas could result in the formation of dense gaseous clumps where dust can accumulate. In a rapid cooling scenario, this could lead to direct gas giant planet formation *(50)*. Considering the gas-dust separation observed in IRS 48, gravitational instability as a cause of the dust concentration is unlikely, since such an instability would concentrate gas and dust simultaneously. Also the gas and temperature properties of the IRS 48 disk are such that gravitational instability is not expected. The classical criteria for gravitational collapse are *(2,51,52)*:

Toomre parameter $Q(r) = \dfrac{c_s(r)\Omega_k(r)}{\pi G \Sigma_{gas}(r)} < 1$

with vertical isothermal sound speed $c_s(r) = \sqrt{\dfrac{k_B T(r)}{\mu m_p}}$

in combination with fast enough cooling time $t_c = \dfrac{\Sigma_{gas}(r) c_s(r)}{2\sigma T^4(r)} < 3\Omega_k(r)^{-1}$

with $k_B$ the Boltzmann constant, $\mu$ the mean molecular mass (2.3), $m_p$ the proton mass, $G$ the gravitational constant and $\sigma$ the Stefan-Boltzmann constant. Applying the formulae to IRS 48 using the gas mass of 27 $M_{Jup}$ in combination with the model temperature structure neither of these requirements are met. Although the assumptions here are simple, the temperature structure or the gas surface density would need to change considerably (more than an order of magnitude) to push the criteria to these limits.

*Planet collision remnants*

If a massive collision happened between two rocky planets, the resulting debris and dust will be first sheared along a ring before it will be turbulently mixed in the radial direction. A single collision is unlikely to have produced the ~9 $M_{Earth}$ of millimeter-sized dust observed here.

*Dust clumps in debris disks*

Concentrations of large grains have been observed in debris disks, where the dust is produced in situ by planetesimal collisions and is then trapped in clumps through resonances with an existing planet *(53,54)*. This scenario is ruled out here by the large amounts of gas still present in the disk outside 25 AU.

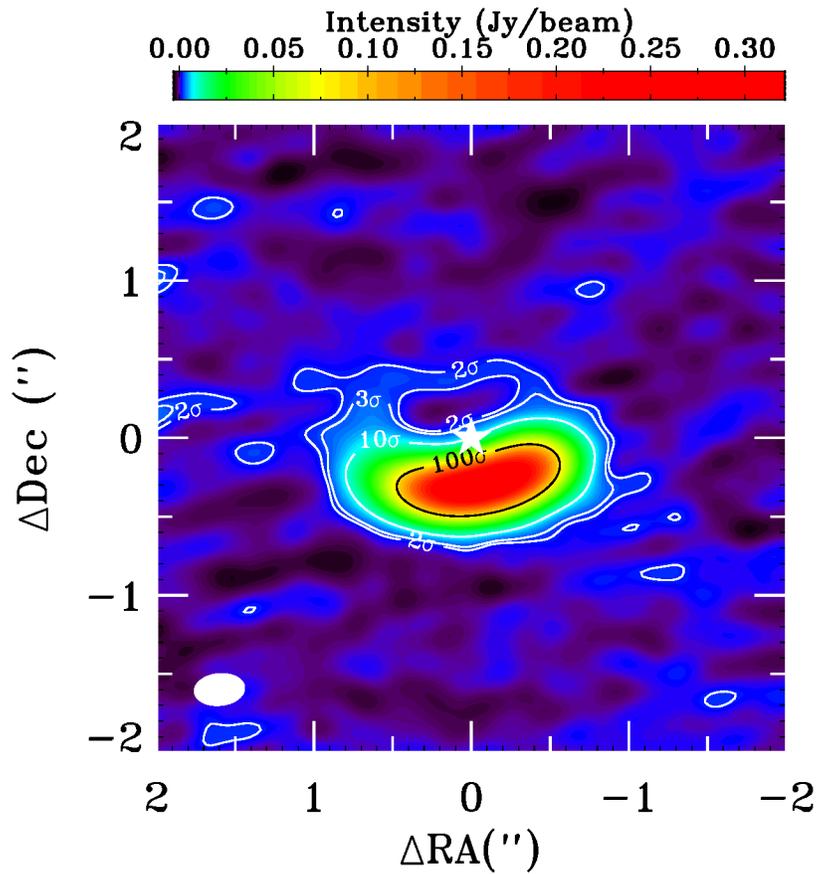

**Fig. S1.** The 0.44 millimeter (685 GHz) continuum emission of IRS 48 as observed with ALMA Band 9 observations, in exponential color scale to show the dynamic range of the image and the high contrast between north and south. Positive values from 0 to 390σ (σ = 0.82 mJy/beam) are shown with blue through red, while negative values from 0 to -4σ are shown with blue, violet, and black. The image is centered on the star (white star symbol). The ALMA beam is indicated with a filled white ellipse in the lower left corner. White and black contours indicate the 2, 3, 10 and 100σ rms levels.

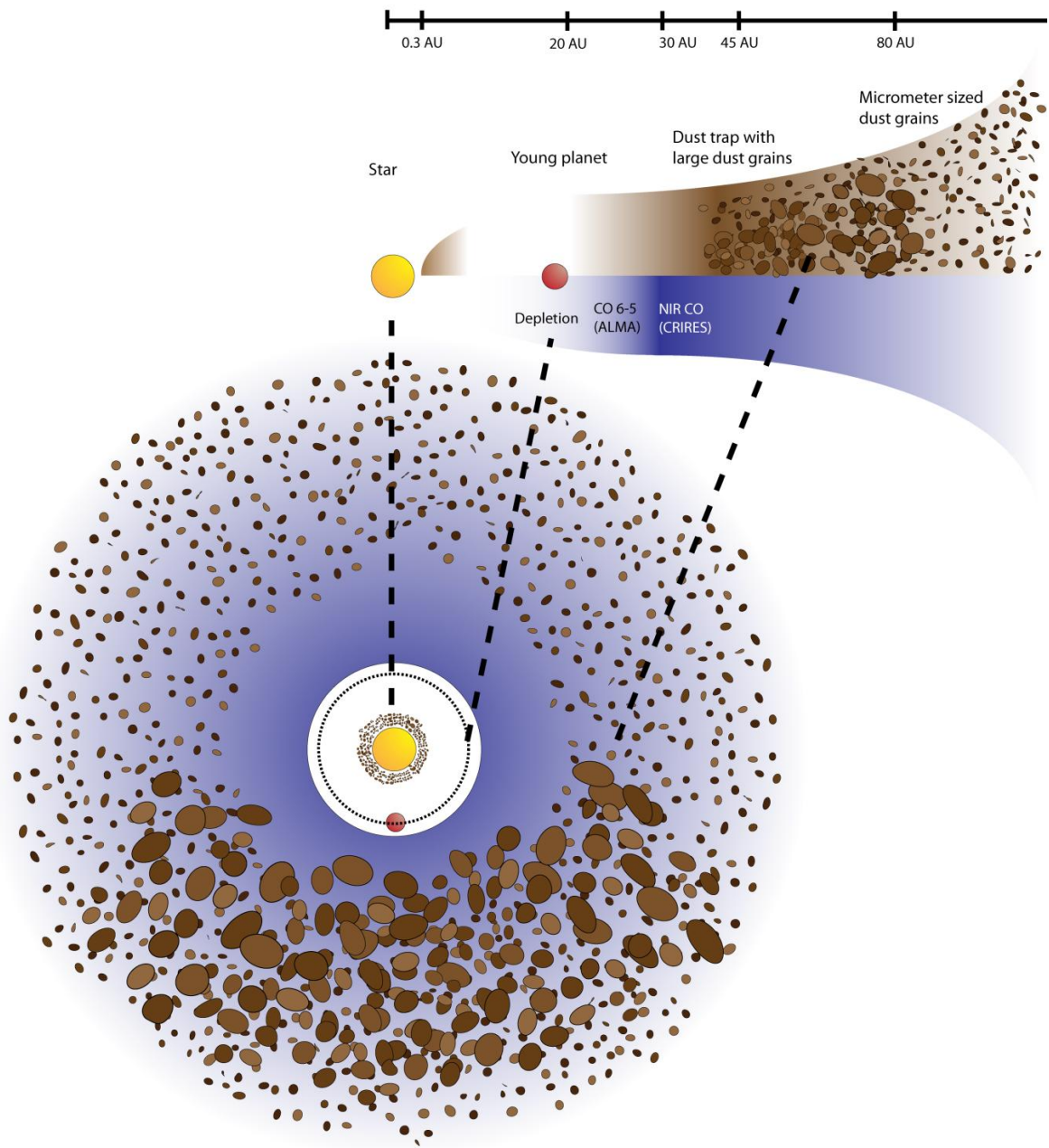

**Fig. S2.** Cartoon of the proposed disk structure of IRS 48. The brown spots represent the large and small grains as traced by the 0.44 millimeter ALMA continuum and VISIR 18.7 micrometer emission, the large grains concentrated in the dust trap in the south. The blue represents the gas surface density as traced by CO 6-5 and fundamental rovibrational CO line as observed by CRIRES, with a gas hole carved out by the planet at 15-20 AU.

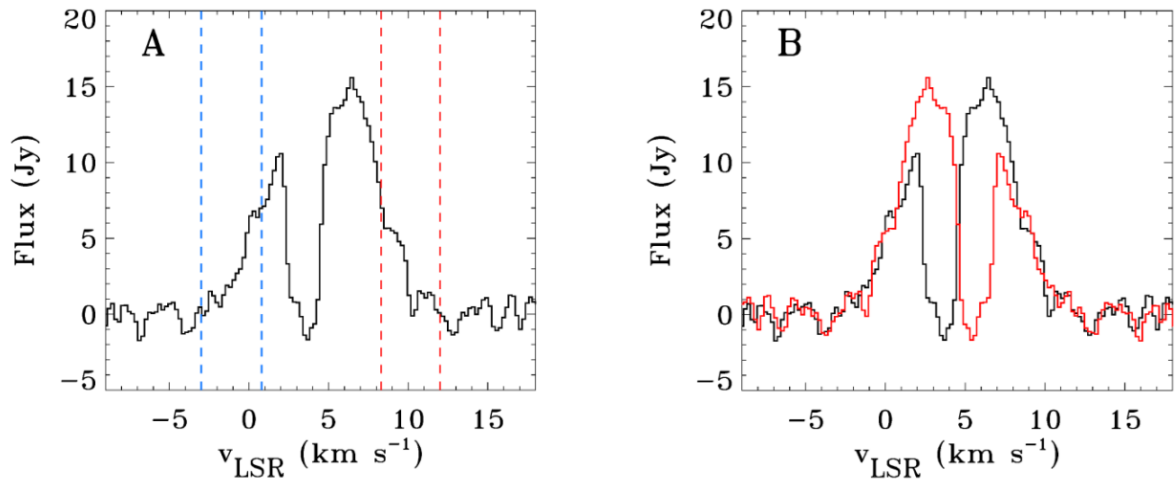

**g. S3.** Full $^{12}$CO 6-5 spectrum integrated over the entire disk area. The profile is distorted by absorption features from foreground clouds around 3.5 km s$^{-1}$. **A)** The velocity limits used for the integrated contour map in Fig. 1 are indicated by blue and red dashed lines for the blue-shifted and red-shifted emission, respectively. The limits are taken at -3, +0.8, +8.3 and +12 km s$^{-1}$. **B)** The spectrum is folded over the source velocity of 4.55 km s$^{-1}$, showing that the spectrum is symmetric, as expected for a Keplerian disk.

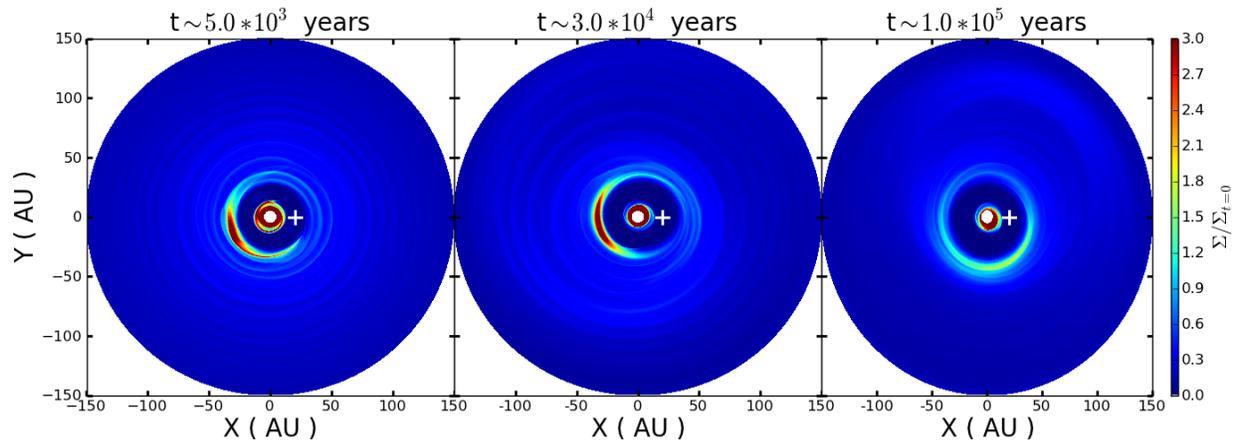

**Fig. S4.** Three snapshots in time of the resulting gas surface density of a disk with a 10 $M_{Jup}$ companion at 20 AU after 1200 companion orbits, simulated with FARGO hydrodynamical code. The companion is indicated with a white cross. The gas surface density was used to model the trapped dust in pressure bumps (Fig. S5B), to simulate the observed dust trap around IRS 48 (Fig. 2).

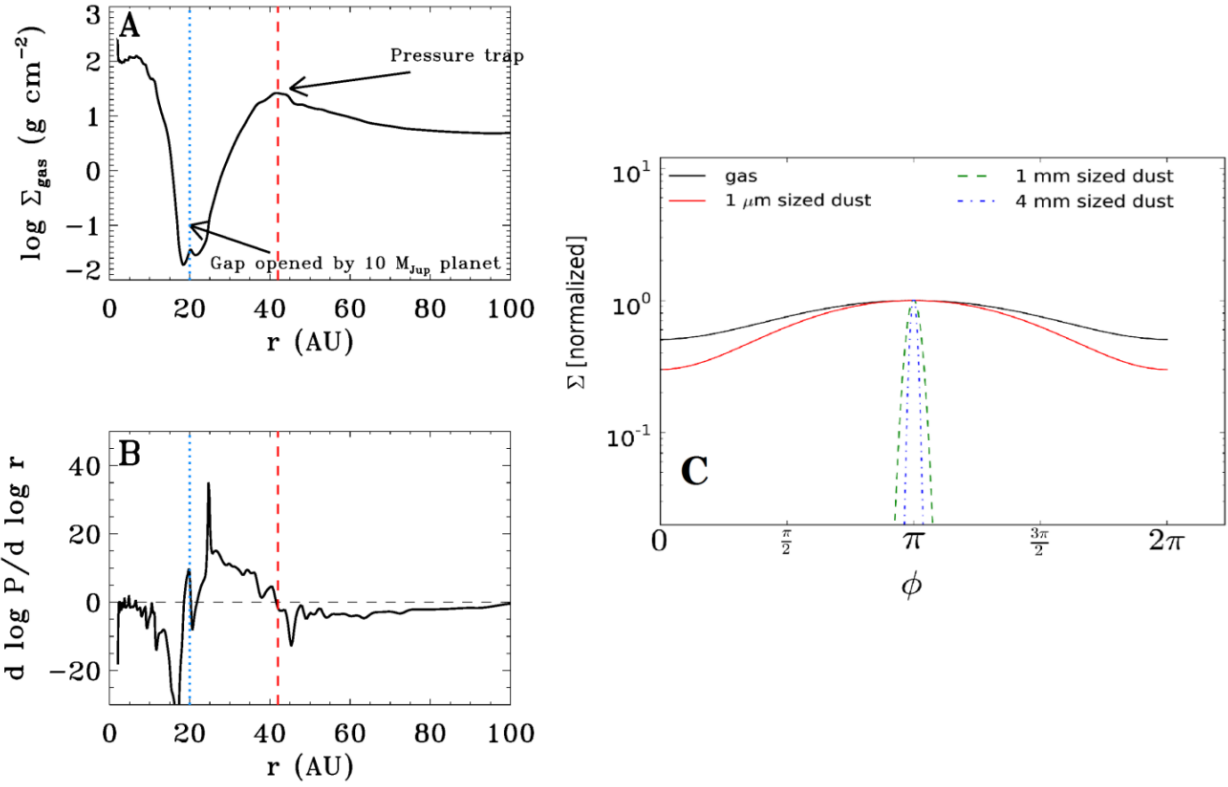

**Fig. S5.** In both A and B, the blue dotted line indicates the location of the planet, the red dashed line the point of zero pressure gradient where the dust gets trapped. **A)** The radial gas surface density $\Sigma_{gas}$ after running >1000 orbits (~0.1 Myr) with a 10 $M_{Jup}$ planet at 20 AU in the model disk. **B)** The pressure gradient $d\log P/d\log r$ corresponding to the gas surface density profile in A). The grey dashed horizontal line shows where the pressure gradient equals zero. **C)** The azimuthal profile of the gas density (black line) versus the large dust grain density (blue) and small dust grain density (red) in the model *(9)*. The contrast in the large dust is significantly higher than in the gas.